\documentclass[a4paper,11pt]{article}
\usepackage{graphicx}

\def\dotdeg{\hbox{$^\circ$\hskip-3pt .}}

\begin{document}
\title{Synchronous Satellites of Venus}
\maketitle\

\begin{center}
\author{Anthony R. Dobrovolskis \\ SETI Institute \\ MS 245-3, NASA Ames Research Center, Moffett Field CA 94035-1000}

Corresponding author:  anthony.r.dobrovolskis@nasa.gov

\vspace{11pt}
\author{Jos\'e Luis Alvarellos \\ Emergent Space \\ MS 202, NASA Ames Research Center, Moffett Field CA 94035-1000}

\end{center}

\vspace{1.0in}

\begin{itemize}
\item 36 pages 
\item 12 figures
\item 5 tables
\end{itemize}

\clearpage

Proposed Running Head: 


Synchronous Satellites of Venus

\vspace{0.5in}
			HIGHLIGHTS


- synchronous satellites of Venus are stable, contrary to conventional lore

- we describe synchronous orbits of Venus

- we discuss applications for synchronous satellites of Venus

\vspace{0.5in}

\begin{abstract}


Synchronous satellites of Venus have long been thought unstable, 
but we use Poincar\'e's surface of section technique to show that 
synchronous quasi-satellites orbiting just outside Venus' Hill sphere 
are quite stable, at least for centuries.  
Such synchrosats always remain within a few degrees of Venus' equator, 
and drift very slowly in longitude.  
These synchrosats could be useful for continuous monitoring 
of points on Venus' surface, such as active landforms or long-lived landers.  

\end{abstract}

\vspace{0.5in}

\textbf{Keywords:} Venus, satellites, orbit, dynamics

\clearpage

\section{Introduction}


The second planet, Venus, is similar to our own planet in terms of mass and size. 
In contrast, however, Venus has a massive atmosphere composed mostly of CO$_2$, 
and is shrouded by a thick layer of clouds, 
which blow around the planet with a period of about four days.  
This cloudy atmosphere causes a runaway greenhouse effect, 
leading to surface temperatures of $\sim$740 K (Hunten, 1999).  

The planet is quite different from ours in another way:  
it is a slow, retrograde rotator, with obliquity of $\sim177.3^\circ$ 
and an inertial rotation rate of one revolution every 243.0 Earth days (see Table 1 for Venusian parameters).  
Adding the latter to its orbital mean motion of one revolution every 224.7 Earth days 
yields the length of Venus' solar ``day'': one diurnal cycle every 116.8 Earth days.  

[INSERT TABLE 1]

\subsection{Satellites of Venus}

Another difference from the Earth is that Venus has no known moons.  
A recent search using the Baade-Magellan 6.5 meter telescope at Las Campanas observatory 
likely rules out moons larger than a few hundred meters (Sheppard and Trujillo, 2009).  

Many authors have noted that strong gravitational perturbations from the Sun 
place severe limits on the long-term stability of Venus orbiters.  For example, 
Buchar (1969) found that prograde moons of Venus had a limiting semi-major axis of 581,000 km, 
but that retrograde satellites had a larger limiting semi-major axis of 655,000 km.  In contrast, 
Rawal (1986) found a prograde limit of 698,000 km, and a retrograde limit of 1,452,000 km.  
For comparison, Venus’ Hill radius (sometimes called its sphere of influence) is $\sim$1,011,000 km.

Inside the above limits, numerous papers (McCord, 1966, 1968; Counselman, 1973; Burns, 1973; Ward \& Reid, 1973; 
Kumar, 1977; Donnison, 1978) all showed that within the age of the Solar system, tidal friction inside Venus 
would have forced large moons either to escape into heliocentric orbit, or to crash into Venus itself.  
Likewise, several papers (McCord, 1968; Singer, 1970; Malcuit \& Winters, 1995) 
have shown that the tidal crash of a retrograde moon could have caused Venus' slow retrograde spin.  
In an interesting variation, Alemi and Stevenson (2006) suggest that a prograde moon of Venus 
may have escaped, but subsequently impacted Venus, reversing its rotation.  

However, satellites are not the only possible companions for a planet; it also may have ``co-orbital'' companions.  
There are now four recognized types of co-orbital companions:  the best known are Trojan companions, 
which librate about a planet's equilateral Lagrange points L4 and L5 in tadpole-shaped orbits.  
Currently thousands of asteroids are known as Trojan companions 
of Neptune, Uranus, Mars, Earth, and Venus, as well as Jupiter.  

``Horse-shoe'' companions librate in wide arcs about the L3 equilibrium point, 
enclosing both L4 and L5 as well.  Several horse-shoe companions of Earth are known, 
and Saturn's small moons Janus and Epimetheus currently share a horse-shoe orbit.  

``Quasi-satellites'' appear to be orbiting a planet outside of its Hill radius 
in the retrograde direction (contrary to the the planet's orbital motion); 
however, they are really not energetically bound to the planet, but orbiting the Sun in the prograde direction.  
Such orbits also are called Distant Retrograde Orbits (DRO); 
St. Cyr {\it et al.} (2000), Stramacchia {\it et al.} (2016), and Perozzi {\it et al.} (2017) 
provide good descriptions of DRO-type missions associated with Earth.  
Currently several quasi-satellites of Earth, Venus, Neptune, and Pluto are known.  

Furthermore, many asteroids are in hybrid co-orbital resonances, 
which alternate among tadpole, horse-shoe, and quasi-satellite configurations.  
Finally, ``counter-orbital'' resonances are also possible, 
when two objects orbit with the same period, but in nearly opposite directions 
(Dobrovolskis, 2012; Morais \& Namouni, 2013, 2016).  
To date, only one counter-orbital companion of Jupiter has been found (Wiegert {\it et al.}, 2017):  
asteroid 2015 BZ 509, also known as (514107) Ka'epaoka'awela.  

While Venus has no known moons, it does have natural co-orbital companions in high-eccentricity orbits, 
namely the quasi-satellite 2002 VE68 (Mikkola {\it et al.}, 2004), 
the horseshoe/quasi-satellite hybrid 2001 CK32 (Brasser {\it et al.}, 2004), 
the Trojan/horseshoe hybrid 2012 XE133 (de la Fuente Marcos \& de la Fuente Marcos, 2013), 
the Earth-crossing Trojan 2013 ND15 (de la Fuente Marcos \& de la Fuente Marcos, 2014), 
and the possible transient Trojan 2015 WZ12 (de la Fuente Marcos \& de la Fuente Marcos, 2017).

The next section describes the Circular Restricted 3-Body Problem (CR3BP), 
and analyzes its usefulness as a model for a satellite of Venus affected by the Sun's gravity; 
while Section 3 applies Poincar\'e's surface of section technique to understand global satellite dynamics.  
Then Section 4 describes periodic orbits around Venus in general, 
while Section 5 describes a unique Venus-synchronous orbit.  
Finally, Section 6 summarizes our findings, and discusses possible applications.

\section{The Circular Restricted 3-Body Problem}

Perturbed two-body dynamics is adequate to describe the behavior of co-orbitals, 
but for a fuller understanding we must turn to the CR3BP, 
where the behavior of a body of infinitesimal mass is affected by the gravitational force 
of two massive bodies orbiting each other.  For our purposes, the two massive bodies are the Sun and Venus, 
and the test particle can represent a co-orbital, a quasi-satellite, or even a Venus orbiter/satellite.

\subsection{Dynamical Description}

The CR3BP is well described in most Celestial Mechanics/Astrodynamics textbooks 
({\it e.g.}, Moulton, 1914; Blanco and McCuskey, 1961; Kaplan, 1976; Danby, 1992; Murray and Dermott, 1999).  
However, in this work we will follow Szebeheley (1967), which the interested reader may consult for details.  
Consider two bodies of masses $m_1$ and $m_2$ revolving about their common center of mass 
in circular orbits due to their mutual attraction, and define the dimensionless mass ratio 
\begin{equation}
			\mu \equiv \frac{m_2}{m_1+m_2} . 
\end{equation}

			We choose the unit of mass such that $m_1 + m_2 = 1$; 
then $m_1 = 1-\mu$ and $m_2 = \mu$.  The distance between $m_1$ and $m_2$ is the unit of length.  
Finally the unit of time is such that the gravitational constant $G$ is unity as well; then by Kepler's third law, 
the mutual orbital period $P$ of $m_1$ and $m_2$ becomes $2\pi$.  Now consider a third body of infinitesimal mass, 
which does not influence the motion of the two massive bodies, but is influenced by them; 
the study of the CR3BP is really the study of the motion of this particle.  

It is convenient to use a rotating Cartesian coordinate system wherein the coordinates 
of $m_1$ and $m_2$ remain fixed at ($\mu$, 0, 0) and ($\mu-1$, 0, 0), respectively.
Then this synodic frame originates at the center of mass of the system, 
such that the (rotating) $x$-axis connects the two massive bodies, 
and the $z$ axis lies along the direction of the angular velocity vector of their mutual orbit, 
while the (rotating) y-axis completes a right-handed triad $(x,y,z)$.  

The conversion between this synodic $(x,y,z)$ frame and a non-rotating (sidereal) coordinate system $(x',y',z')$ 
with the same origin is then 
\begin{equation}
			x'= x \cos(t) - y \sin(t) ,
\end{equation}
\begin{equation}
			y'= x \sin(t) + y \cos(t) ,
\end{equation}
\begin{equation}
			{\rm and} \; \; z' = z , 
\end{equation}
where $t$ is the time.  Note that the orbital angular velocity of the two massive bodies is unity 
because of our choice of the unit of time.  


In the synodic (rotating) frame, the equations of motion for the test particle are 
\begin{equation}
			\ddot{x} - 2 \dot{y} = \partial U/\partial x ,
\end{equation}
\begin{equation}
			\ddot{y} + 2 \dot{x} = \partial U/\partial y ,
\end{equation}
\begin{equation}
			{\rm and} \; \; \ddot{z} = \partial U/\partial z, 
\end{equation}
					where 
\begin{equation}
			U = x^2/2 + y^2/2 + (1-\mu)/r_1 +\mu/r_2 .  
\end{equation}

Here $r_1$ and $r_2$ are the distances of the test particle from $m_1$ and $m_2$ respectively, and are given by 
\begin{equation}
			r_1 = \sqrt{ (x-\mu)^2 +y^2 +z^2}, 
\end{equation}
\begin{equation}
			r_2 = \sqrt{ (x-\mu +1)^2 +y^2 +z^2} . 
\end{equation}

To within an arbitrary constant, the CR3BP has only one integral of motion, namely the Jacobi constant 
\begin{equation}
			C \equiv 2U - \dot{x}^2 -\dot{y}^2 -\dot{z}^2 +\mu[1-\mu].  
\end{equation}
Note that by these conventions, $U$ is minus the usual potential energy (centrifugal plus gravitational), 
while $C$ is --2 times the usual total energy (potential plus kinetic), plus a constant term $\mu[1-\mu]$.  
This constant is chosen so that $C$ = 3 when $r_1 = r_2$ = 1 and $\dot{x} = \dot{y} = \dot{z}$ = 0.

\subsection{The Lagrange Points}

The three-dimensional CR3BP has three degrees of freedom, and since only one constant of motion exists, 
there is no general solution in closed form to the equations of motion (5) through (7).  
However, specific solutions do exist.  

The equations of motion contain five equilibria, named the {\it Lagrange points}, and labeled L1, L2, L3, L4, and L5.  
The first three are called the collinear Lagrange points because they all lie on the $x$ axis; 
in our coordinates, L2 lies between the primary mass $m_1$ and the secondary mass $m_2$, 
but L1 lies on the side of $m_2$ opposite $m_1$, while L3 lies on the side of $m_1$ opposite $m_2$.  

The two remaining points L4 and L5 are called the triangular Lagrange points, 
because they form equilateral triangles with $m_1$ and $m_2$ in the plane $z = 0$; 
in our coordinates, L5 lies $60^\circ$ ahead of $m_2$ in its orbit about $m_1$, 
while L4 lies $60^\circ$ behind $m_2$.  It can be shown that the collinear points are always unstable, 
but the triangular points are stable provided that $\mu < (1-\sqrt{23/27})/2 \approx$ 0.0385208965 (Szebehely, 1967).

\subsection{The Zero Velocity Curves}

It can be seen from Definition (11) of the Jacobi constant $C$ that there are constraints on the motion 
of the test particle; clearly motion is only possible when $0 < \dot{x}^2 +\dot{y}^2 +\dot{z}^2 = 2U -C +\mu[1-\mu]$.  
Then for a given value of $C$, 
setting $C < 2U +\mu[1-\mu]$ gives Zero Velocity Curves (ZVCs) which define regions where motion is allowed.  

There are good descriptions and sketches of these curves in the aforementioned references, especially Szebehely (1967).  
Very briefly, at high enough values of $C$ the ZVCs form one closed oval around $m_1$, 
and another around $m_2$, as well as a larger outer oval encompassing both masses; 
motion of the test particle is only possible inside the inner smaller ovals or outside the larger one 
(in this situation either $r_1$ or $r_2$ is very small, or $r_0 \equiv \sqrt{x^2+y^2+z^2}$ is fairly large).  

As the value of $C$ is lowered to a critical value $C_2$, the two inner ovals meet at the L2 point; 
and as $C$ is lowered even further, the L2 point provides an avenue through which the particle 
can move between the two inner ovals.  As we lower $C$ even more to another critical value $C_1$, 
the oval surrounding $m_2$ ``touches'' the L1 point; and as $C$ is lowered even further, 
this point provides an opening through which the particle can escape to infinity.  

Continuing with this process, we reach another critical value $C = C_3$ where the ZVCs meet at the L3 point.  
As we lower $C$ even further towards $C_4 = C_5$ = 3, the ZVCs shrink to smaller and smaller ovals 
surrounding the triangular Lagrange points L4 and L5, and vanish completely for $C < 3$.  
Note that the ZVCs are symmetric about the (rotating) $x$ axis and that, 
the lower the Jacobi constant, the more space is available for the test particle to explore.

\subsection{Applicability to the Sun-Venus-satellite case}

The CR3BP assumes that the two massive bodies orbit each other in perfectly circular orbits.  
Venus orbits the Sun in a slightly elliptical orbit; however, Venus' orbital eccentricity 
$e_V \approx 0.000677$ is the smallest of any planet in the Solar System (Murray and Dermott, 1999), 
so to first order in $e_V$ we are justified in neglecting its eccentricity.  

A second effect we ignore is that Venus is not perfectly spherical.  
Its dynamical oblateness dominates solar perturbations for orbits inside 
about 2.6 Venus radii (Capderou, 2005).  
Another perturbation we neglect is the gravity of the other planets.  
We find that the perturbation of the Earth on satellites of Venus is roughly four orders of ten smaller 
than the solar perturbation.  The other planets have even smaller influences on Venus satellites.  

A massive satellite also would be subject to tidal perturbations.  
We estimate that even a close moon of Venus would have to have a mass on the order 
of $3 \times 10^{17}$ kg (about 30 times the mass of Mars' larger moon Phobos) 
for tides to be as important as solar perturbations; 
for more remote satellites, this mass increases as the eighth power of the distance.  
Therefore we ignore tidal perturbations as well.  

[INSERT TABLE 2]

\section{Surfaces of Section}


The solution to the CR3BP may be viewed as a flow in a six-dimensional phase space.  
However, the system may be confined to only four dimensions by setting $z$ and $\dot{z}$ to zero 
in Eqs. (5) through (7); thus motion takes place in the $(x,y)$ plane.  
Then this becomes the so-called Plane Circular Restricted 3-Body Problem.  Furthermore, 
in the Plane CR3BP the existence of the Jacobi constant constrains the phase space motion to take place 
on a three-dimensional manifold.  We now consider the intersection of a two-dimensional plane with this manifold.  

Following Smith and Szebehely (1992) and Alvarellos (1996), we choose $y=0$, {\it i.e.} the $(x,\dot{x})$ plane; 
furthermore we investigate only the positive crossings ($\dot{y} > 0$).  
This is the concept of the Poincar\'e surface of section (Lichtenberg and Lieberman, 1992).  
We used many different initial conditions to construct our surfaces of section, 
and wrote a program to propagate each initial condition forward in time 
with a fifth-order Runge-Kutta (Cash-Karp) variable step-size integrator (Press {\it et al.}, 1992).

Figure 1 shows a surface of section for $C$ = 3.0010, 
a relatively high value of the Jacobi constant compared to $C_1$ and $C_2$ (see Table 2), 
so that the Zero Velocity Curves around Venus are closed.  
In this plot the location of Venus is approximately (--1.0, 0.0); 
on either side of the vertical line at $x \approx -1$, 
we can see sets of concentric curves known as ``invariant curves'' (Lichtenberg and Lieberman, 1992).  
Each of these curves represents a quasi-periodic orbit, 
while the point at the center of each family of concentric curves 
represents a simply periodic orbit, a topic which we will revisit later.  

[INSERT FIGURE 1:  C = 3.0010]

In general, those concentric curves on the right-hand side of Fig. 1 ($x > -1$) represent prograde orbits; 
these objects orbit Venus in the same direction as Venus orbits the Sun. Those on the left ($x < -1$) 
represent retrograde orbits, which orbit Venus in the opposite direction as Venus orbits the Sun.  
The dash-dot curves enveloping all orbits represent the projection of the ZVC 
on the $(x,\dot{x})$ plane (H\'enon, 1970).  
For this relatively high value of the Jacobi constant, objects revolve about Venus in small orbits, close to the planet.  
Note that we see a somewhat random scattering of points near the central line $x = -1$; 
these points represent chaotic orbits, so here we see the onset of chaos.  






For comparison, Fig. 2 plots a surface of section for the critical value $C_1$ = 3.000776... (Table 2).  
For this value, the ZVCs around Venus are just barely closed at the L1 point, located at (--1.00937..., 0); 
but the ZVCs are open at the L2 point, at (--0.99068..., 0), so that satellites can escape to inner heliocentric orbits.  
Indeed this is an effective mechanism for Venus to lose prograde satellites, 
since chaotic orbits tend to ``explore'' all of the phase space available to them (Lichtenberg and Lieberman, 1992).  

[INSERT FIGURE 2:  C = C1 = 3.000776...]

Finally, Fig. 3 plots a surface of section for $C$ = 3.0006, 
a relatively low value of the Jacobi constant relative to $C_1$ and $C_2$, 
so that the ZVC around Venus are now wide open at both the L1 and L2 points.  
Note how empty the right side of the plot is, 
as most prograde orbits have turned chaotic and escaped to heliocentric space; 
but most retrograde orbits still show signs of stability, although we see some hints of developing chaos.  
These surfaces of section demonstrate the well-known fact that retrograde orbits are more stable than prograde ones.  

[INSERT FIGURE 3:  C= 3.0006]

\section{Periodic Orbits}

As mentioned in the previous section, as the invariant curves in the surface of section plots shrink down to a point, 
that point itself represents a (simply) periodic orbit.   A periodic orbit has the property that 
$x$, $y$, $z$, $\dot{x}$, $\dot{y}$, and $\dot{z}$ at time $t$ are all the same as at time $t +T$, 
where $T$ is the period of that orbit.  


Note that $T$ represents the period only in a synodic frame of reference rotating with period $P$; in a non-rotating 
(sidereal) frame, a prograde (or ``direct'') orbit completes one revolution during a time interval $T_D$, where 
\begin{equation}
					1/T_D = 1/T +1/P . 
\end{equation}
In contrast, a retrograde orbit of period $T$ in a synodic frame completes 
one revolution in a non-rotating frame during a time interval $|T_R|$, where 
\begin{equation}
					1/T_R = 1/T -1/P . 
\end{equation}




In what follows we search for periodic orbits about the secondary mass, namely Venus. 
We have obtained our periodic orbit solutions via a manual/iterative surface-of-section method.  
A more rigorous approach would be to use continuation methods, such as those described by Davis (1962).

\subsection{Prograde Periodic Orbits (g and g' families)}

Inspection of Figure 1 reveals the concentric invariant curves on both sides, and from the right-hand side   
we can obtain approximate initial conditions for a simply-periodic prograde orbit: $x(0) \approx$ --0.997 and $\dot{x}(0)$ = 0 (this 
is where the curves shrink down to a point).  From the definition of the surface of section, $y(0)$ 
vanishes, while the value of $\dot{y}(0)$ can be obtained from Equation 11.  Using this information
we manually iterate until we find the periodic orbit (i.e., the trajectory exactly repeats).
The ``measured'' period $T$ of this orbit is about 0.318; recall that the orbital period of Venus around 
the Sun is 224.7 days, equivalent to $2\pi$, from which we get $T$ = 11.4 days.  
This prograde, periodic orbit around the secondary body (Venus in our case) 
belongs to the g-family of orbits, which originate as close prograde satellite orbits (H\'enon 1969).

The same procedure allows us to obtain initial conditions for other prograde periodic orbits 
corresponding to the aforementioned surface of section plots.  
As we move from high to low values of $C$ and reach the critical value $C_2$, 
an interesting thing occurs:  the prograde periodic orbits ``split''.  
In other words, for a given value of $C < C_2$ we find \textit{two} simply-periodic orbits rather than just one.  
We see hints of this in the two islands floating in the chaotic region along the x-axis seen on the right-hand side of Fig. 2.  
This bifurcation phenomenon was already noted by H\'enon (1969), and he assigned these prograde orbits to the g'-family.  
There is a unique correspondence between a given value of $C$ and a unique g-family prograde periodic orbit, 
while this is not true for the g'-family (H\'enon, 1969).  
Detailed initial conditions for these orbits are shown in Table 3.

[INSERT TABLE 3]

Figure 4 plots the g and g' families of orbits around Venus from Table 3, with orbital periods in the range 11.4 $\le T \le$ 60.7 days.  
For comparison, the dotted circle represents Venus' Hill sphere, with radius 
\begin{equation}
                                                R_H = (\mu/3)^{1/3} \approx 0.00934 ; 
\end{equation}
while each of the Lagrange points L1 and L2 is marked with an X.  
Note that the distance from Venus to either L1 or L2 is approximately equal to its Hill radius.  

[INSERT FIGURE 4] 

The innermost g orbit in Fig. 4 corresponds to a Jacobi constant $C$ = 3.0015.  To obtain smaller, tighter orbits we would 
need to go to even higher values of $C$.  Such orbits would resemble Keplerian orbits as $C$ increases still further.  

At the other end of the spectrum, we can obtain larger periodic g' orbits at lower Jacobi constant values; 
these become even more distorted until we reach a collision orbit.  
Thereafter more prograde periodic orbits are found until again 
we reach a second collision orbit, and so on (Szebehely 1967; H\'enon 1969).

\subsection{Retrograde Periodic Orbits (f family)}

By applying the same approach from the previous section to the retrograde orbits, 
we obtain the initial conditions shown in Table 4.  
Figures 5 through 7 display the orbits thus obtained; 
these represent members of the f family of periodic orbits, 
which originate as close retrograde satellite orbits (H\'enon, 1969).  

[INSERT TABLE 4]

Figure 5 shows the inner periodic f orbits; note that in contrast to the prograde case, 
there is no splitting, and the orbits shown here are very close to circular.  
Again, the orbits become small and nearly Keplerian for high values of $C$.  
Surfaces of section are useful to obtain initial conditions, 
but they are not strictly necessary: by carefully extrapolating from existing periodic orbits we can obtain further 
periodic orbits without them.  

[INSERT FIG. 5] 


Figure 6 shows larger periodic f orbits; as in Fig. 4, the dotted circle represents Venus' Hill sphere, 
for comparison, while again each of the Lagrange points L1 and L2 is marked with an X.  
Note that the orbit labeled 8 is comparable in size to Venus' Hill sphere.  
At some point between the orbits labeled 7 and 8, trajectories become noticeably non-circular, 
with their $y$ diameter growing larger than their $x$ diameter.  

[INSERT FIG. 6] 
 

Finally, Fig. 7 shows the evolution of even larger retrograde periodic f orbits.  
As we lower the value of $C$, trajectories become even more distorted and kidney-shaped.  
Note how orbit 13 crosses the dashed circle of radius 1.3825 representing the Earth's orbit around the Sun, 
as well as the dotted circle of radius 0.5352 representing Mercury's orbital semi-major axis; 
such an orbit cannot be expected to be stable in the real Solar system.  
The same applies to orbit 14; eventually we reach orbit 15, which collides with the Sun.  
This only represents the end of the first phase of retrograde periodic orbits; 
beyond the collision orbit we can find further periodic orbits that loop around the Sun 
and return to ``orbit'' the whole system (Szebehely, 1967).  

[INSERT FIG. 7]

\section{Synchronous Satellites}

Now reconsider the otherwise undistinguished retrograde f orbit number 9, just outside Venus' Hill sphere in Fig. 6.  
Table 4 shows that its period $T$ in a synodic frame is 116.8 days, 
while its period $T_R$ in a non-rotating (sidereal) frame is 243 days.  
These are just the same as Venus' rotation period in those frames; 
thus a particle in this orbit would appear to hover almost stationary over Venus' surface, 
like a synchronous satellite.

\subsection{Dynamical Considerations}

Figure 8 plots both the orbit of Venus and the synchronous orbit in a sidereal frame with the Sun at the origin.  
To produce Fig. 8, we compute the particle's position in the sidereal frame via Eqs. (2) and (3) (we confine ourselves 
to the $xy$ plane), while the Sun's sidereal position as a function of time is given by ($\mu \cos t$, $\mu \sin t$); 
then we determine the relative position of the particle with respect to the Sun, obtaining 
\begin{equation}
			x'_{\rm rel} = (x-\mu) \cos(t) - y \sin(t) 
\end{equation}
			and
\begin{equation}
			y'_{\rm rel} = y \cos(t) + (x-\mu) \sin(t) .
\end{equation}

[INSERT FIGURE 8] 


In Fig. 8a we see the two orbits are very close at this scale.  Starting at the leftmost position at time $t$ = 0, 
the particle is farther away from the Sun than Venus (Fig. 8b); but moving counter-clockwise, at $t = P/4$ (56 days) 
the particle is now closer to the Sun (Fig. 8c).  The pattern then repeats, since at $P/2$ (112 days) 
the particle is again farther away from the Sun; while at $3P/4$ (169 days), the particle is closer again.  

The particle's distance $r_2$ from the center of Venus 
ranges from $\sim1.2 \times 10^6$ km to $\sim1.6 \times 10^6$ km ($\sim200$ to $\sim260$ Venus radii), 
with an oscillation period of only $T/2 \approx$ 58.3 day.  In contrast, 
the particle's distance $r_1$ from the Sun varies between 107.0 $\times 10^6$ and 109.4 $\times 10^6$ km, 
but with an average value approximately equal to Venus' semi-major axis (108.2 $\times 10^6$ km), 
and an oscillation period $T$ of $\sim$116.8 days.  For comparison, 
Venus' Hill radius $R_H \approx$ 1.011 $\times 10^6$ km in physical units.  



The particle's sidereal speed $V$ is 
\begin{equation}
			V^2 = ( dx'_{\rm rel}/dt )^2 + ( dy'_{\rm rel}/dt )^2 
\end{equation}
\[
    = (x^2+y^2) + (\dot{x}^2+\dot{y}^2) + 2(x\dot{y}-y\dot{x}) - 2\mu(x+\dot{y}) + \mu^2 .
\]
Converting from CR3BP units to physical units, we find that $V$ varies between 34.4 and 35.6 km/s 
for a synchronous satellite of Venus, with an average value approximately equal to Venus' orbital speed (35.0 km/s); 
the period of these oscillations is again $T \approx 116.8$ days.  

The particle's heliocentric semi-major axis $a$ can be obtained from the vis-viva equation 
\begin{equation}
			\frac{(1-\mu)}{2a} = \frac{1-\mu}{r_1} -V^2/2 
\end{equation}
(Danby, 1992).  For a synchronous satellite of Venus, $a$ varies between 106.6 and 109.9 $\times 10^6$ km;
its average value is approximately equal to Venus' semi-major axis, while its period of oscillations is again $T$.  

To obtain the particle's heliocentric eccentricity $e$, 
we also need the orbit's specific angular momentum $h$ with respect to the Sun.  
Given the heliocentric position (Eqs. 15 and 16) and velocity, after some algebra we obtain
\begin{equation}
			h = r^2_1 + (x-\mu)\dot{y} - y \dot{x} . 
\end{equation}
Then the eccentricity $e$ can be obtained from 
\begin{equation}
				h^2 = a(1-e^2) 
\end{equation}
(Danby, 1992).  For a synchronous satellite of Venus, $e$ varies between 0.0204 and 0.0267, 
with an average of $\sim$0.0237 and an oscillation period of just $T/2$.  
Table 5 summarizes the characteristics of this synchronous orbit.  



[INSERT TABLE 5]



Figure 9 shows orbit number 9 again, but this time in a non-rotating (sidereal) frame centered on Venus.  
In this frame, note how the quasi-satellite circulates slowly clockwise about Venus, in a nearly square trajectory.  
For comparison, Fig. 10 shows orbit number 9 once more, but this time in a rotating frame centered on Venus, 
and fixed in its body.  In this frame, 
the particle circulates counter-clockwise twice during each $T$ = 116.8-day period around an apparent oval path 
of longitudinal (east-west) radius $\sim2.5 \times 10^5$ km, and a radius of $\sim1.6 \times 10^5$ km 
in the radial direction, centered at a point fixed in this frame $\sim1.4 \times 10^6$ km from the center of Venus.  

[INSERT FIG. 9 - SYNCHROSAT IN NON-ROTATING FRAME]

[INSERT FIG. 10 - SYNCHROSAT IN BODY FRAME]

Thus a spacecraft in this orbit would appear to hover indefinitely above a nearly fixed point on Venus' equator, 
like a synchronous satellite, while librating in longitude by $\sim \pm11^\circ$, with a period of 58.4 days.  
Its libration in latitude due to Venus' obliquity is only $\sim \pm3^\circ$, with a period of 243 days.  


Synchronous satellites have long been considered impossible for both Mercury and Venus, 
because they were thought to be unstable ({\it e.g.}, Capderou, p. 467), 
unlike synchronous satellites of the Earth or Mars.  
A straightforward application of Kepler's Third Law to synchronous orbits around Venus 
gives a semi-major axis of 1,530,500 km (Wertz, 2011).  
This lies outside the usual stability limit of $\sim R_H/2$,
corresponding to an orbital period $p$ of only $\sim$0.35 times 
the planet's heliocentric orbital period $P$ (Hamilton \& Burns, 1991).  

However, the above limit applies only to {\it prograde} satellites.  
We have seen from the surface of  section plots that retrograde orbits are more stable than prograde ones.  
Alvarellos (1996) found the same stability limit as Hamilton and Burns (1991) did for prograde satellites, 
while he found a limit of $\sim R_H$ for retrograde satellites {\it in circular orbits}.  

When the assumption of circular orbits is relaxed, however
{\it there is no formal outer limit to retrograde satellites} 
({\it e.g.}, Jackson, 1913; H\'enon, 1969, 1970; Benest, 1971, 1979).
Specifically, the f family of retrograde, periodic orbits are very stable; 
indeed, Figs. 5 through 7 indicate very large, retrograde trajectories.  
Henon and Guyot (1970) state that if $0 \leq \mu \leq$ 0.0477 
(a condition certainly met by the Sun-Venus system), 
then ``... essentially all retrograde periodic orbits around the lighter $m_2$ body are stable''.

Thus Mercury, with a prograde obliquity of only $0.033^\circ$, an orbital period $P$ of 87.97 days, and a long 
rotation period of 58.65 days = $2P/3$, cannot retain synchronous satellites against the tidal pull of the Sun.  
In contrast, because Venus is a slow {\it retrograde} rotator, it can retain synchronous quasi-satellites indefinitely. 

Of course, a spacecraft in a synchronous orbit is not really affected by Venus' rotation; 
its synodic period just happens to match the planet's diurnal period.  
The principal effect of Venus' gravity is just to keep the object's epicenter 
(the center of its epicycle in the synodic frame) librating about Venus' location.  


\subsection{Stability Considerations}


We have tested the stability of this venusynchronous orbit with the freely-available NASA software 
called the General Mission Analysis Tool, or GMAT (version R2018a)\footnote{https://sourceforge.net/projects/gmat/}.  
Advantages of using GMAT as a propagator are that it is very easy 
(a) to switch central bodies; (b) to turn forces on or off; and (c) to find the spacecraft groundtrack.  

To begin, we included only the point-mass gravitational forces of the Sun and Venus.  
We transformed the initial conditions of orbit 9 from Table 4 to a heliocentric, 
Cartesian state vector in the J2000 frame at its Epoch (1-Jan-2000, 12:00:00.0 UTC), 
and we used the Runge-Kutta 89 integrator to propagate them forward in time for 10 Earth years.  
The resulting orbit is co-planar with Venus' orbit, and the putative spacecraft 
always remains just outside Venus' Hill radius for the duration of the integration. 
Over this time period, the sub-satellite latitude stays within $\sim2.7^\circ$ of Venus' equator, 
while its longitude spans less than $60^\circ$.  Over a full rotation of Venus (243 days), 
the sub-satellite longitude spans only $\sim26^\circ$.

Next we took the GMAT-baseline case and turned on the different individual forces one by one, 
to investigate their effects on the groundtrack.  
We found that solar radiation pressure has negligible effects on a ten-year time-scale, 
using values typical for a geosynchronous telecommunications satellite: 
a Solar Radiation Pressure (SRP) coefficient of 1.8, 
along with an area-to-mass ratio value of 0.04 m$^2$/kg.  


The gravitational effects of Mercury and Mars were negligible as well; however, 
we found that the gravitational effect of the Earth is to drift the groundtrack noticeably eastward, 
while the gravitational effect of Jupiter is to drift the groundtrack in the opposite direction.  

Figure 11 plots this groundtrack, taking into account all realistic forces (point gravitational effects 
of the Sun, Mercury, Venus, Earth, Mars, Jupiter, Saturn, Uranus and Neptune, as well as SRP); 
their overall effect is to drift the pattern westward while the sub-satellite latitude always remains near the equator.
The groundtrack drifts westward, regardless of the starting Epoch; 
if it is deemed necessary to keep the groundtrack constrained to a certain small region, 
the spacecraft may be equipped with a propulsion system to counteract the main perturbative effects 
({\it i.e.}, Earth and Jupiter).  Finally, Fig. 12 plots the evolution of the heliocentric orbital elements 
with all relevant forces previously mentioned turned on; the orbit is highly stable.  

[INSERT FIGURE 11]

[INSERT FIGURE 12]

\section{Discussion/Conclusions}

We investigated co-orbital companions of the planet Venus using the planar CR3BP as a model, 
with the Sun and Venus as the massive bodies.  We used surfaces of section 
to find an interesting orbit whose sidereal period matches Venus' rotational period of 243 days; 
an object in this orbit would in principle remain nearly stationary above a point on the equator of the planet, 
akin to a stationary orbiter.  In practice the groundtrack is not strictly stationary, 
but oscillates about a fixed longitude on the equator of Venus (it also oscillates slightly in latitude).  
This orbit lies beyond Venus` Hill radius, and a spacecraft located there is more properly said to orbit the Sun, 
and not Venus itself.  

Since this particular distant retrograde orbit (DRO) is stable in the CR3BP, 
we also investigated its stability in a more realistic ephemeris model with all additional relevant forces included.  
We have also found the orbit to be stable in the more realistic model.  
Note that a simple application of Kepler's Third Law to compute the distance to a stationary satellite of Venus 
also gives a (different) value beyond the Hill radius, 
but placing a spacecraft in a circular orbit at that distance is not viable 
because the satellite would be quickly lost to heliocentric space.  
The only way to have a spacecraft in a stable venusynchronous orbit 
is to place it in the specific orbit which we have found in this paper; 
it is a heliocentric orbit located in Venus' orbital plane, perturbed by the gravity of Venus itself.  

Venusynchronous satellites would be rather far away from the planet; 
from such a vantage point, Venus' angular diameter ranges from $0.4^\circ$ to $0.6^\circ$, 
about as big as our Moon looks from Earth ($0.5^\circ$).  
In contrast, the Earth spans a diameter of $17.4^\circ$ from a geostationary satellite (Soop, 1994).  
Despite this distance, one prime advantage of venusynchronous satellites 
is that of continuous monitoring of the full planetary disk at a nearly fixed longitude; 
there are several other advantages as well (Montabone {\it et al.}, 2020).  

Venusynchronous satellites could be equipped with appropriate imaging equipment, 
since Venus' surface is invisible to telescopic eyes and its cloudy atmosphere 
is transparent in only a few spectral windows in the near-infrared (Hunten, 1999).  
They could also be used to maintain constant contact for long intervals with long-lived landers, 
such as those NASA is trying to develop (Kremic, Hunter, \& Tolbert, 2019); 
or continously to monitor specific surface features, such as Idunn Mons, 
a young volcano located at $46^\circ$ south latitude in the Imdr Regio volcanic province (D'Incecco {\it et al.}, 2017), 
especially given recent reports of possible present-day volcanism on Venus 
(Filiberto {\it et al.}, 2020; G\"{u}lcher {\it et al.}, 2020; see also Esposito, 1984 and Marcq {\it et al.}, 2013).

 
One advantage of venusynchronous satellites over geostationary ones is that of power: 
in the vicinity of the Earth, $\sim$1.4 kW/m$^2$ of solar flux are available for spacecraft operations (Soop, 1994); 
but for Venus spacecraft, the equivalent value is $\sim$2.6 kW/m$^2$, almost twice as much.  
For the latter, and assuming that the synchrosat bus is similar to geostationary ones, 
the solar array drive assemblies (SADAs) need to complete a full rotation in 116.8 days to track the Sun.  
However, these Venus synchrosats would experience more challenging thermal issues than the typical Earth satellite.  

Geosynchronous satellites require batteries to store energy for periodic intervals in Earth's shadow; 
they undergo two eclipse seasons centered around each equinox, with a maximum eclipse duration of about 72 minutes.  
In addition, operators need to account for less-frequent but occasionally deep lunar eclipses.  
In contrast, Venus synchrosats would never be in total eclipse, because Venus' umbra 
(total shadow, of length = $a_V R_V/(R_{Sun} -R_V) \approx$ 949000 km) lies inside its Hill sphere.  

Because of perturbations, geostationary spacecraft typically need to perform periodic maneuvers 
to maintain their orbital slots, {\it i.e.}, ``stationkeeping'' (Soop, 1994).  
An uncontrolled geosynchronous satellite will oscillate back and forth in longitude 
around the nearest of the two stable equilibrium points with a period of more than two years.  
The main perturbations affecting mission operations are luni-solar gravities (affecting mainly the spacecraft's 
inclination); the asymmetric geopotential, the main terms being $J_2$ (oblateness, affecting inclination) 
and $J_{22}$ (equatorial ellipticity, affecting the longitude); and SRP (affecting the eccentricity).  

The dynamical environment for a putative venusynchronous satellite is quite different, though.  
Since such a spacecraft is far away from Venus, the $J_2$ and $J_{22}$ perturbations would be minimal.  
Regarding luni-solar perturbations, of course Venus has no moon, and the Sun-Venus arrangement 
is in fact what keeps the spacecraft in its synchronous orbit; hence there is no need to correct for these effects.  
The only significant perturbation a spacecraft in this orbit may need to correct for are the effects of SRP 
on the eccentricity.  Note that SRP is strongly dependent on the spacecraft's area-to-mass ratio; 
therefore the smaller this value, the less propellant will be needed.

Orbit insertion requirements for a venusynchronous satellite 
also should be much less demanding than those for a close satellite.  
A small Venus synchrosat might be dropped off by a vehicle on its way to Venus, 
such as NASA's VERITAS and DAVINCI+ missions, or ESA's EnVision mission, 
or by one flying past Venus {\it en route} to another destination.  

For comparison, a retrograde satellite of Venus with a period of only 4 days 
(synchronous with the ``super-rotation'' of Venus' cloud layer) could be useful 
for long-term monitoring of cloud features, trace gases, or a long-lived balloon.  
A simple calculation indicates that such a satellite would have an orbital 
semi-major axis of $\sim100,000$ km $\approx$ 16.5 Venus radii, 
much more comparable to synchronous satellites of Earth or Mars.

\section{Acknowledgments}

We would like to thank L. Dones for some references, as well the GMAT, Tex Maker, and GNU Octave teams.  
Thanks also to L. Plice, N. Alem and E. Tapio.  JLA thanks his family for their patience 
(Alejandra; Jose Jr; Isabella and Daniel; my `pichus').  Dedicated to the memories of Y. Geremia and N.E.P.  
JLA also would like to thank Ravi Mathur of Emergent Space for previewing the manuscsript; 
we also thank Giovanni Valsecchi and an anonymous referee for reviewing it.

\section{References}

\setlength{\parindent}{-0.2in}

Alemi, A., and D. Stevenson (2006).  Why Venus has no moon (abstract).  {\it B.A.A.S.} {\bf 38}, 491.  

Alvarellos, J. L. (1996).  Orbital Stability of Distant Satellites of Jovian Planets.  
M.Sc. thesis, San Jose State University.  

Benest, D. (1971).  Elliptic restricted problem for Sun-Jupiter:  
Existence of stable retrograde satellites at large distance.  
{\it Astronomy \& Astrophysics} {\bf 13}, 157--160.  

Benest, D. (1979).  Large satellite orbits of Jupiter 
and Earth in the perturbed restricted three-body problem.  
Pp. 225--230 in {\it Natural and Artificial Satellite Motion}, 
ed. P. Nacozy and S. Ferraz-Mello.  U. of Texas Press, Austin.  

Blanco V. M. and S. W. McCuskey (1961).  Basic Physics of the Solar System.  
Addison-Wesley Publishing Co., Massachussetts.  

Brasser, K., K. A. Innanen, M. Connors, C. Veillet, P. Weigert, S. Mikkola, and P. W. Chodas (2004).  
Transient co-orbital asteroids.  {\it Icarus} {\bf 171}, 102-109.  

Buchar, E. (1969).  A short study on the motion of artificial satellites 
of Mercury and Venus (abstract).  P. 41 in COSPAR-IAU-IAG/IUGG-IUTAM 
Symposium {\it Dynamics of Satellites}, Prague; ed. B. Morando.  

Burns, J. A. (1973).  Where are the satellites of the inner planets ?  
{\it Nature Physical Science} {\bf 242}, 23--25.  

Capderou, M. (2005).  Satellites: Orbits and Missions.  Springer-Verlag, France.  

Counselman, C. C. III (1973).  Outcomes of tidal evolution.  {\it Astrophysical J.} {\bf 180}, 307--314.  

Danby, J. M. A. (1992).  Fundamentals of Celestial Mechanics.  Willman-Bell, Inc.  Richmond.  

Davis, H. T. (1962).  Introduction to Nonlinear Differential and Integral Equations.  Dover, New York.  

de la Fuente Marcos, C., and R. de la Fuente Marcos (2013).  
Asteroid 2012 VE133:  A transient companion to Venus.  {\it M.N.R.A.S.} {\bf 432}, 886--893.  

de la Fuente Marcos, C., and R. de la Fuente Marcos (2014).  
Asteroid 2013 ND15:  Trojan companion to Venus, PHA to the Earth.  
{\it M.N.R.A.S.} {\bf 439}, 2970--2977.  

de la Fuente Marcos, C., and R. de la Fuente Marcos (2017).  
Transient co-orbitals of Venus:  An update.  
{\it Research Notes of the AAS} {\bf 1:1}.  

D'Incecco, P., N. M\"uller, J. Helbert, M. D'Amore (2017).  
Idunn Mons on Venus: Location and extent of recently active lava flows.  
{\it Planet. Space Sci.} {\bf 136}, 25–-33.  

Dobrovolskis, A. R. (2012). Counter-orbitals: another class of co-orbitals.
{\it AAS Division for Planetary Sciences Meeting} {\bf 44}, abstract 112.22 .  

Donnison, J. R. (1978).  The escape of natural satellites from Mercury and Venus.  
{\it Astrophysics and Space Science} {\bf 59}, 499--501.  

Esposito, L. W. (1984).  Sulfur Dioxide:  Episodic injection shows evidence for active Venus volcanism.  
{\it Science} {\bf 223}, 1072--1074.  

Filiberto, J., D. Trang, A. H. Treiman, and M. Gilmore (2020).  Present-day volcanism on Venus 
as evidenced from weathering rates of olivine.  {\it Science Advances} {\bf 6}, 1--4.  

G\"{u}lcher, A. J. P.,  T. V. Gerya, L. G. J. Montési and J. Munch (2020).  
Corona structures driven by plume-lithosphere interactions and evidence 
for ongoing plume activity on Venus. {\it Nature Geoscience} {\bf 13}, 547--554.  

Hamilton, D. P., and J. A. Burns (1991). Orbital stability zones about asteroids. {\it Icarus} {\bf 92}, 118--131.  

H\'enon, M. (1969).  Numerical exploration of the restricted problem, V.  
Hill's case: periodic orbits and their stability.  {\it Astronomy and Astrophysics} {\bf 1}, 223--238.  

H\'enon, M. (1970).  Numerical exploration of the restricted problem, VI.  
Hill's case: non-periodic orbits.  {\it Astronomy and Astrophysics} {\bf 9}, 24--36.  

H\'enon, M., Guyot, M. (1970).  Stability of periodic orbits in the restricted problem.  
In Periodic Orbits, Stability and Resonances, ed. G. E. O. Giacaglia.  
Dodrecht-Holland: D. Reidel Publishing Company.  

Hunten, D. (1999).  Venus: Atmosphere.  Pp. 147--159, in Encyclopedia of the Solar System, 
ed. P. R. Weissman, L. McFadden, and T. V. Johnson.  Academic Press.  

Jackson, J. (1913).  Retrograde satellite orbits.  {\it M.N.R.A.S.} {\bf 74}, 62--82.  

Kaplan, M. H. (1976).  Modern Spacecraft Dynamics and Control.  Wiley and Sons, New York.  

Kremic, T., G. Hunter, and C. Tolbert (2019).  Status of developments 
toward long-lived Venus landers.  {\it EPSC Abstracts} {\bf 13}, 2068.  

Kumar, S. S. (1977).  The escape of natural satellites from Mercury and Venus.  
{\it Astrophysics and Space Science} {\bf 51}, 235--238.  

Lichtenberg, A. J., and M. A. Lieberman (1992).  Regular and Chaotic Dynamics (2d Ed).  Springer-Verlag, New York.  

Malcuit, R. J., and R. R. Winters (1995).  Numerical simulation of retrograde gravitational capture 
of a satellite by Venus:  Implications for the thermal history of the planet (abstract).  
{\it L.P.S.C.} {\bf 26}, 885--886.  

Marcq, E., J.-L. Bertaux, F. Montmessin, and D. Belyaev (2013).  
Variations of sulphur dioxide at the cloud top of Venus's dynamic atmosphere.  
{\it Nature Geoscience} {\bf 6}, 25--28.  

McCord, T. B. (1966).  Dynamical evolution of the Neptunian system.  {\it Astronomical J.} {\bf 71}, 585--590.  

McCord, T. B. (1968).  The loss of retrograde satellites in the solar system.
{\it J. Geophysical Research} {\bf 73}, 1497--1500.  

Mikkola, S., R. Brasser, P. Wiegert, and K. Innanen (2004).  
Asteroid 2002 VE 68, a quasi-satellite of Venus.  {\it M.N.R.A.S.} {\bf 351}, L63--L65.  

Montabone, L., N. Heavens, and 38 co-authors (2020).  Observing Mars from areostationary orbit:  
benefits and applications.  White paper submitted to the Planetary Science and Astrobiology Decadal Survey 2023--2032.  

Morais, M. H. N., and F. Namouni (2013).  
Retrograde resonance in the planar three-body problem.  {\it Cel. Mech. Dyn. Astron.} 125, 91–106.  

Morais, M. H. N., and F. Namouni (2016).  
A numerical investigation of coorbital stability and libration in three dimensions.  
{\it Cel. Mech. Dyn. Astron.} {\bf 125}, 91–106.  

Moulton, F. R. (1914).  An Introduction to Celestial Mechanics (2d Ed).  
MacMillan.  Reprinted in 1970 by Dover, New York.  

Murray, C. D., and S. F. Dermott (1999).  Solar System Dynamics.  Cambridge U. Press, Cambridge.  

Perozzi, E., M. Ceccaroni, G. B. Valsecchi, and A. Rossi (2017).  
Distant retrograde orbits and the asteroid hazard.  {\it The European Physical Journal Plus} {\bf 132}, 1--9.  

Press, W., {\it et al.} (1992).  Numerical Recipes in FORTRAN.  Cambridge U. Press, Cambridge.  

Rawal, J. J. (1986).  Possible satellites of Mercury and Venus.  
{\it Earth, Moon, and Planets} {\bf 36}, 135--138.  

Sheppard, S. S., and C. A. Trujillo (2009).  A survey for satellites of Venus.  {\it Icarus} {\bf 202}, 12--16.  

Singer, S. F. (1970).  How did Venus lose its angular momentum?  {\it Science} {\bf 170}, 1196--1198.  

Sjogren, W. L., W. B. Banerdt, P. W. Chodas, A. S. Knopliv, G. Balmino, J. P. Barriot, J. Arkani-Hamed, 
T. R. Colvin, and M. E. Davis (1997).  The Venus gravity field and other geodetic parameters.  
Pp. 1125--1161 in Venus II, ed. S. W. Bougher, D. M. Hunten, and R. J. Phillips.  U. of Arizona Press.  

Smith, R. H., and V. Szebehely (1992).  
The onset of chaotic motion in the restricted problem of three bodies.  
{\it Cel. Mech. Dyn. Astron.} {\bf 56}, 409--425.  

Soop, E. M. (1994).  Handbook of Geostationary Orbits.  Springer/Kluwer.  

St. Cyr, O. C., M. A. Mesarch, H. M. Maldonado, D. C. Folta, A. D. Harper, J. M. Davila, and R. R. Fisher (2000).  
Space Weather Diamond: a four spacecraft monitoring system.  
{\it J. of Atmospheric and Solar-Terrestrial Physics} {\bf 62}, 1251--1255.  

Standish, E. M. (1998). JPL Planetary and Lunar Ephemerides, DE405/LE405, JPL IOM 312.F-98-048.  

Stramacchia, M., C. Colombo, and F. Bernelli-Zazzera (2016).  
Distant retrograde orbits for space-based Near Earth Objects detection.  
{\it Advances in Space Research} {\bf 58}, 967--988.  

Szebehely, V. G. (1967).  Theory of Orbits.  Academic Press, New York.  

Ward, W. R., and M. J. Reid (1973).  Solar tidal friction and satellite loss.  {\it M.N.R.A.S.} {\bf 164}, 21--32.  

Wertz, J. R. (2011).  Orbits and Astrodynamics.  
Pp. 197--234 in Space Mission Engineering: The New SMAD, 
eds. J. R. Wertz, D. F. Everett, and J. J. Puschell.  Microcosm Press.  

Wiegert, P., M. Connors, and C. Veillet (2017).  
A retrograde co-orbital asteroid of Jupiter.  
{\it Nature} {\bf 543}, 687–689.

\vspace*{0.2in}

\begin{quotation}
\noindent
Table 1. Properties of Venus.
\end{quotation}

\begin{center}
\begin{tabular}{|c|c|c|c|}
\hline
Parameter        	&	Symbol	&	Value				&	Source			\\
\hline
			&		&	1.082089 $\times 10^8$ km	&				\\
Semi-major axis		    &	$a_V$	&	(0.72333199 AU)       		&    Murray \& Dermott (1999)	\\
Eccentricity		    &	$e_V$	&	0.00677323			&	 "			\\
Obliquity               &            	&	$177.3^\circ$			&	 "			\\
Inclination to Ecliptic	&		&	$3\dotdeg39$                    & 				\\
Gravitational Parameter	&	$Gm_p$	&	324858.601			&				\\
			&		&	$\pm$ 0.014 km$^3$/s$^2$	&  Sjogren {\it et al.} (1997)	\\
Equatorial Radius	    &	$R_p$	&	6051.0 km                       &	 "			\\
Hill Radius	        &	$R_H$	&	1011000 km			&				\\
Orbital Period		    &	$P$	&	224.7 days                      &				\\
Rotation Period		    &           &	243.0 days                      &				\\
\hline
\end{tabular}
\end{center}


\newpage

\noindent
Table 2. Sun-Venus CR3BP Parameters.  The value of $\mu$ is computed using the $GM$ values for Venus (see Table 1) 
and the Sun (1.3271244002 $\times 10^{11}$ km$^3$/s$^2$ Standish, 1998).  
Locations of the Lagrange points are given in the synodic (rotating) frame.  

\begin{center}
\begin{tabular}{|c|c|c|}
\hline
	Parameter			&	Value         			& Page ref. in Szebehely (1967)	\\
\hline
	Mass Parameter $\mu$		&   2.44783236410728 $\times 10^{-6}$	&		p. 217		\\
\hline
	Location of the L1 point	&	(-1.0093710166, 0)              &		"		\\
	Location of the L2 point	&	(-0.9906822994, 0)              &		p. 221		\\
	Location of the L3 point	&	(+1.0000010199, 0)              &		p. 225		\\
  Location of the L4 point$^{\dagger}$	&   (-0.4999975522, +0.8660254038)	&   				\\
  Location of the L5 point$^{\dagger}$	&   (-0.4999975522, -0.8660254038)	&   				\\
\hline
		$C_1$			&		3.0007768995		&		p. 217		\\
		$C_2$			&		3.0007801633		&		p. 221		\\
		$C_3$			&		3.0000048957		&		p. 225		\\
		$C_4$			&		3			&				\\
		$C_5$			&		3			&				\\
\hline
\end{tabular}
\end{center}
{$\dagger$ The locations of the L4 and L5 points are ($\mu -0.5$, $\pm\sqrt{3}/2$). }

\vspace*{0.2in}

\noindent
Table 3. Prograde periodic orbits around Venus: Initial conditions in CR3BP units 
(in addition note that $y(0) = z(0) = \dot{x}(0) =\dot{z}(0) = 0$).  
$T$ and $T_D$ are the periods in the synodic and sidereal frames, respectively.  
Note that orbits 1-3 have unique Jacobi constants (g-family), 
but orbits 4 and 5 have the same value of the Jacobi constant (g'-family); see Fig. 4.  

\begin{center}
\begin{tabular}{|c|c||c|c||c|c|c||c|}
\hline
		&		&			&			&	$T$	&	$T$	&	$T_D$	&			\\
	Orbit	&	$C$	&	$x(0)$		&	$\dot{y}(0)$	&	(CR3BP)	&	(days)	&	(days)	&	Notes		\\
\hline
	1	&	3.0015	&   -0.998229599...	&	0.035653318...	&	0.318	&	11.4	&	10.8	&                   	\\
	2	&	3.0010  &   -0.997092625...	&	0.026520158...	&	0.718	&	25.7	&	23.0	&  Fig. 1$^{\dagger}$	\\
	3	&	3.0009  &   -0.996670046...	&	0.024438066...	&	0.926	&	33.1	&	28.9	&                       \\
\hline
	4	&	$C_2$   &   -0.998216072...	&	0.044385974...	&	1.636	&	58.5	&	46.4	&                       \\
	5	&	$C_2$   &   -0.993100000...	&	0.008102525...	&	1.698	&	60.7	&	47.8	&	             		\\
\hline
\end{tabular}
\end{center}
{$\dagger$~Right side of the surface of section.}\\

\newpage


\noindent

Table 4. Retrograde periodic orbits around Venus: Initial Conditions in CR3BP units 
(in addition note that $y(0) = z(0) = \dot{x}(0) =\dot{z}(0) = 0$).  
These represent the f-family of periodic orbits.  
$T$ and $T_R$ are the periods in the synodic and sidereal frames, respectively.  
Orbits 1--6 are shown in Fig. 5, Orbits 6--10 in Fig. 6, and Orbits 7--15 in Fig. 7.  

\begin{center}
\begin{tabular}{|c|c||c|c||c|c|c||c|}
\hline
		&		&			&			&	$T$	&	$T$	&	$T_R$	&			\\
	Orbit	&	$C$	&	$x(0)$		&	$\dot{y}(0)$	&	(CR3BP)	&	(days)	&	(days)	&	Notes		\\
\hline
	1	&	3.0015	&   -1.001497449...	&	0.041992835...	&	0.226	&	8.08	&	8.38	&  			\\
	2	&	3.0010	&   -1.002120439...	&	0.036225577...	&	0.374	&	13.37	&	14.22	&  Fig. 1$^{\dagger}$	\\
	3	&	3.0009	&   -1.002312066...	&	0.034984347...	&	0.422	&	15.09	&	16.18	&			\\
	4	&	$C_2$	&   -1.002580804...	&	0.033580263...	&	0.494	&	17.67	&	19.17	&			\\
	5	&	$C_1$	&   -1.002580048...	&	0.033636896...	&	0.496	&	17.74	&	19.26	&  Fig. 2$^{\dagger}$	\\
	6	&	3.0006	&   -1.003114808...	&	0.031500033...	&	0.646	&	23.10	&	25.75	&  Fig. 3$^{\dagger}$	\\
\hline
	7	&  3.0002677...	&	-1.0050000	&	0.0297900	&	1.212	&	43.3	&	53.7	&			\\
	8	&  2.9999879...	&	-1.0090000	&	0.0281100	&	2.554	&	91.3	&	153.9	&			\\
	9	&  2.9999046...	&	-1.0111475	&	0.0299500	&	3.266	&	116.8	&	243.	&	See Text	\\
	10	&  2.9997765...	&	-1.0150000	&	0.0348000	&	4.328	&	154.8	&	498.	&			\\
\hline
	11	&  2.9974926...	&	-1.0500000	&	0.0993000	&	6.190	&	221.37	&	14922	&			\\
	12	&  2.9596138...	&	-1.2000000	&	0.3835000	&	6.274	&	224.37	&	153142	&			\\
	13	&  2.7323333...	&	-1.5000000	&	0.9225000 	&	6.276	&	224.44	&	195710	&			\\
	14	&  2.3284617...	&	-1.7500000	&	1.3700000	&	6.282	&	224.66	&	1170804	&            		\\
	15	&  1.0597786...	&	-2.0000000	&	1.9850000	&	$2\pi$	&	224.70	&	$\infty$&	Collision	\\
	16	& -1.6662078...	&	-2.2500000	&	2.7600000	&	12.560	&	449.17	&	-449.63	&	Circum-		\\
		&		&			&			&		&		&		&	binary		\\
\hline
\end{tabular}
\end{center}
{$\dagger$~Left side of the surface of section.} \\

\noindent
Table 5. Venusynchronous orbit characteristics.  
The heliocentric orbital elements are specified in a J2000 Ecliptic frame for 01-Jan-2000, 12:00:00 UTC (J2000 Epoch).
Specific initial conditions in CR3BP units and other characteristics are shown in Table 4, orbit 9.  

\begin{center}
\begin{tabular}{|l|c|c|c}
\hline
Parameter                                        & Value                               & Notes            \\
\hline
Semi-major axis (km)                             & 106590220.95                        &                  \\
Eccentricity                                     & 0.022717                            &                  \\
Inclination (deg)                                &   3.39471                           & Same as Venus    \\
Longitude of Asc. Node (deg)                     &  76.68069                           &      "           \\
Argument of Perihelion (deg)                     & 298.94917                           &                  \\
True Anomaly (deg)                               & 166.95154                           &                  \\
\hline
Average heliocentric semi-major axis, $a$        & 1.0823 $\times 10^8$ km (0.7235 AU) & See Fig. 12a     \\               
Average heliocentric eccentricity, $e$           & 0.0237                              & See Fig. 12b     \\
\hline
Average distance from Venus                      & 1.414 $\times 10^6$ km              & See Fig. 10     \\
\hline

\end{tabular}
\end{center}

\clearpage

\section{Figure Captions}

\begin{description}


\item[Figure 1] Surface of section for the Jacobi constant value $C$ = 3.0010; 
the dashed curves represent the zero velocity curves (ZVCs).  


 
\item[Figure 2] Surface of section for the Jacobi constant value $C = C_1$ (3.000776...).  
The L1 point is located at $x \approx$ --1.0094, $dx/dt$ = 0, where the ZVCs meet.  
The Sun (not shown) is located to the right at $(\mu, 0)$.  
Note the widespread evidence of chaos on the right-side of the plot (prograde orbits).
A particle in a prograde orbit is free to escape the vicinity of Venus towards the Sun through the L2 point.  

\item[Figure 3] Surface of section for the Jacobi constant value $C$ = 3.0006 .  Most prograde orbits 
have turned chaotic and escaped the vicinity of Venus to escape either to the Sun through the L2 point, 
or to infinity through the L1 point.  In contrast, retrograde orbits stay in the vicinity of Venus.  

\item[Figure 4] Prograde periodic orbits around Venus (shown as the black dot; not to scale) in the synodic frame.  
The direction of motion is counterclockwise.  The numbers correspond to the initial conditions shown in Table 3.  
Orbits 1--3 belong to the g-family, while orbits 4 and 5 belong to the g'-family (left and right branches, respectively).  
The dotted circle represents Venus' Hill sphere, while the Lagrange points L1 and L2 are marked with crosses.  
The Sun (not shown) is located to the right at $(\mu, 0)$.  

\item[Figure 5] Inner retrograde periodic orbits around Venus (shown as the black dot; not to scale) 
in the synodic frame.  The direction of motion is clockwise.  The numbers correspond to the initial conditions 
shown in Table 4.  Orbits 4 and 5 are indistinguishable at this scale.  
The Sun (not shown) is located to the right at $(\mu, 0)$.  

\item[Figure 6] Middle retrograde periodic orbits around Venus (shown as the black dot; not to scale) in the synodic 
frame.  The direction of motion is clockwise.  The numbers correspond to the initial conditions shown in Table 4.  
The dotted circle represents Venus' Hill sphere, while the Lagrange points L1 and L2 are marked with crosses.  The Sun 
(not shown) is located to the right at $(\mu, 0)$.  For scale, orbit 6 is shown in both this and the previous figure.  

\item[Figure 7] Outer retrograde periodic orbits around Venus (shown as the black dot; not to scale) 
in the synodic frame.  The direction of motion is clockwise.  
The numbers correspond to the initial conditions shown in Table 4.  The outermost orbit shown is a collision orbit.  
The locations of the Lagrange points L3, L4 and L5 are marked with crosses.  The Sun is located at $(\mu, 0)$ (shown as 
the yellow dot; not to scale).  For comparison, the dashed circle represents the orbit of the Earth around the Sun, 
while the dotted circle represents the orbital semi-major axis of Mercury.  

\item[Figure 8] (a) Orbits of Venus (solid) and its synchronous satellite (dotted; orbit 9 in Figure 6/Table 4) 
in a non-rotating (sidereal) frame centered at the Sun.  The length scale is in km.  
On this scale, the orbits of Venus and its synchrosat are almost indistinguishable; 
for context, the ellipse represents the orbit of Mercury.  
In this frame, all three bodies move counter-clockwise; 
locations of Venus and its synchrosat are shown at four different times, in days.  
(b) Close-up at $t$ = 0; the synchrosat is farther from the Sun than Venus.  
(c) Close-up at $t$ = 56.2 days; the synchrosat is closer to the the Sun than Venus.  



\item[Figure 9] Orbit 9 viewed in a non-rotating (sidereal) frame centered on Venus, over an interval of 715 days.  
In this frame, the Sun completes a revolution ``around'' Venus in 224.7 days.  
The dotted circle represents Venus' Hill sphere.  

\item[Figure 10] Orbit 9 viewed in a reference frame centered on Venus, and fixed in its body.  
The polar angle represents longitude as measured on the surface of Venus.  
The dotted circle represents Venus' Hill sphere. 



\item[Figure 11] Orbit 9 groundtrack on a Venus cloud background:  10-year simulation performed with GMAT 
(SRP taken into account, as well as gravitational forces of the Sun and all the planets).  
The groundtrack shifts westwards with time.  The inset shows the groundtrack in detail.  

\item[Figure 12] Evolution of the heliocentric orbital elements of a putative venusynchrosat over 10 years 
(GMAT simulations with all relevant forces included).  The starting time corresponds to 1-Jan-2000, 12:00:00 UTC, 
and the reference frame is J2000 Ecliptic (see Table 5).  The orbit is stable.


\end{description}

\clearpage

\clearpage
\begin{figure}
\begin{center}
\includegraphics[width=160mm]{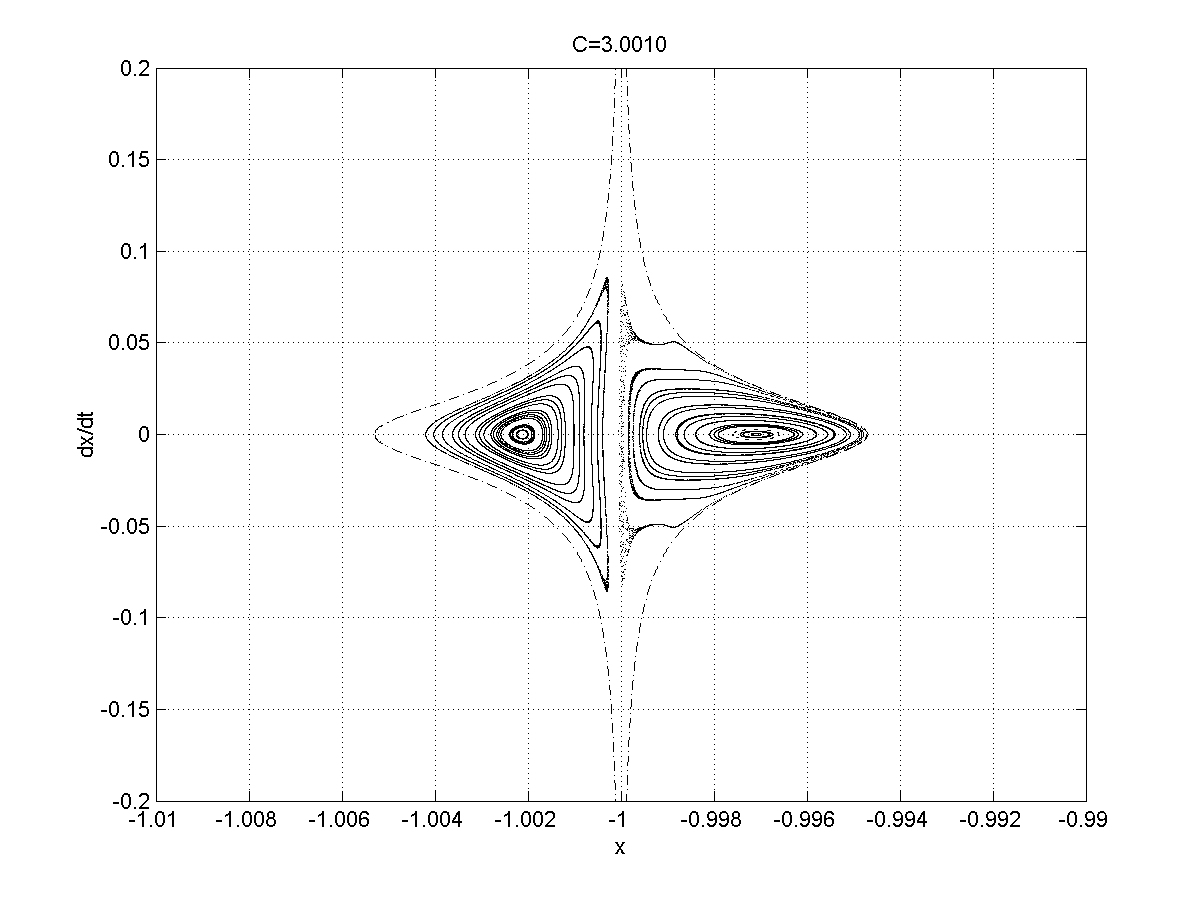}
\caption{}
\end{center}
\end{figure}

\clearpage
\begin{figure}
\begin{center}
\includegraphics[width=160mm]{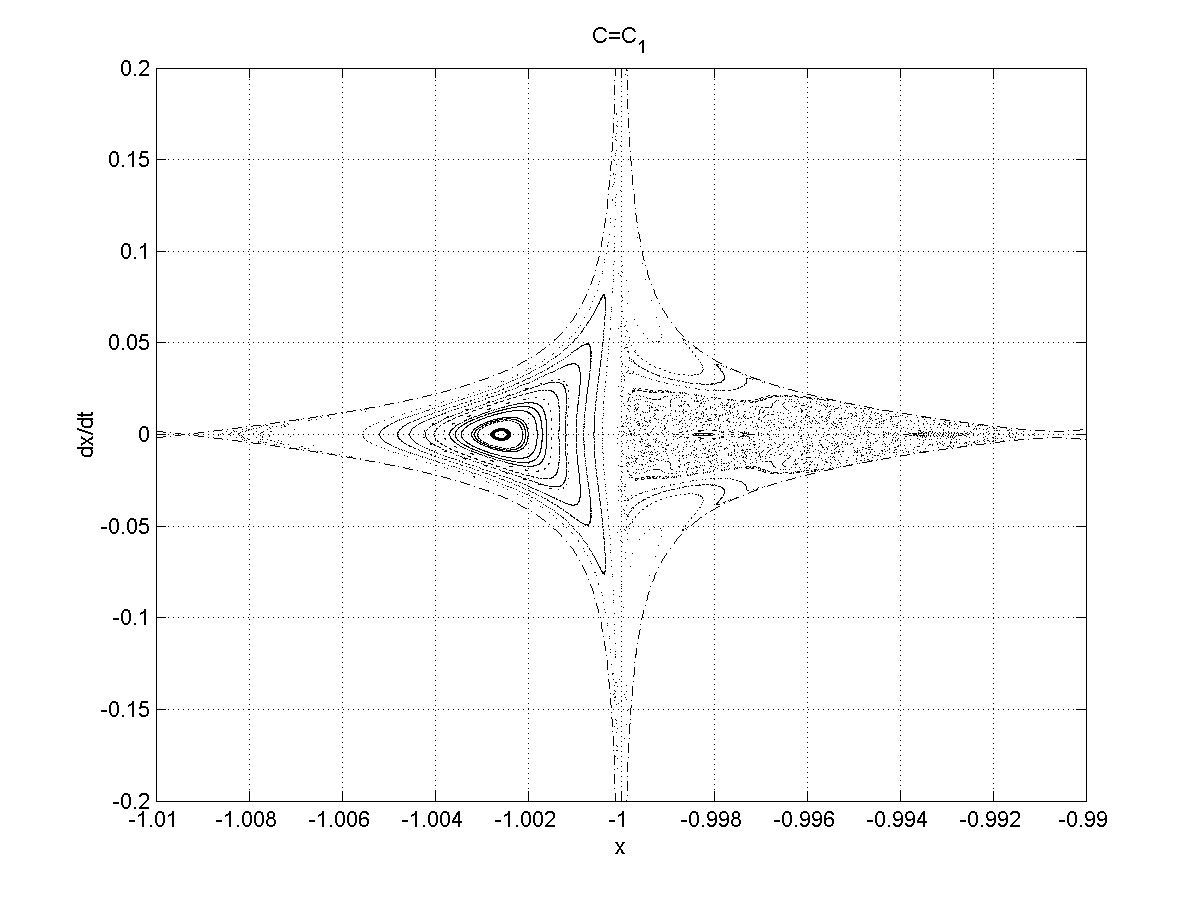}
\caption{}
\end{center}
\end{figure}

\clearpage
\begin{figure}
\begin{center}
\includegraphics[width=160mm]{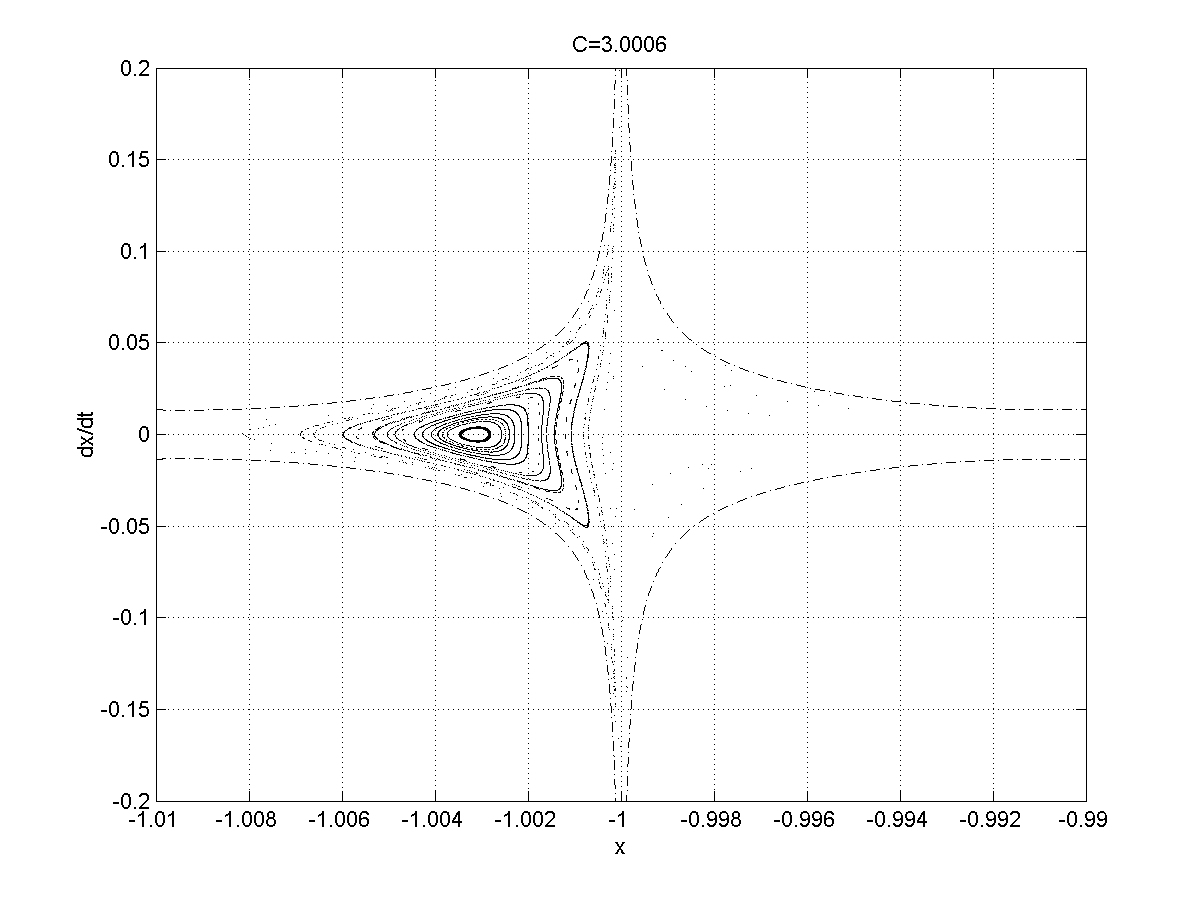}
\caption{}
\end{center}
\end{figure}
 
\clearpage
\begin{figure}
\begin{center}
\includegraphics[width=160mm]{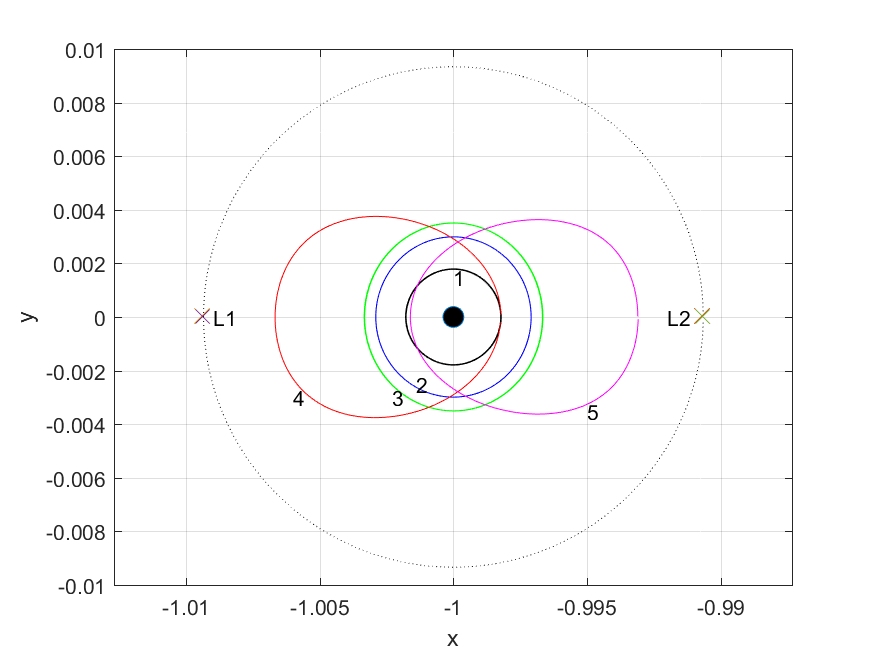}
\caption{}
\end{center}
\end{figure}

\clearpage
\begin{figure}
\begin{center}
\includegraphics[width=160mm]{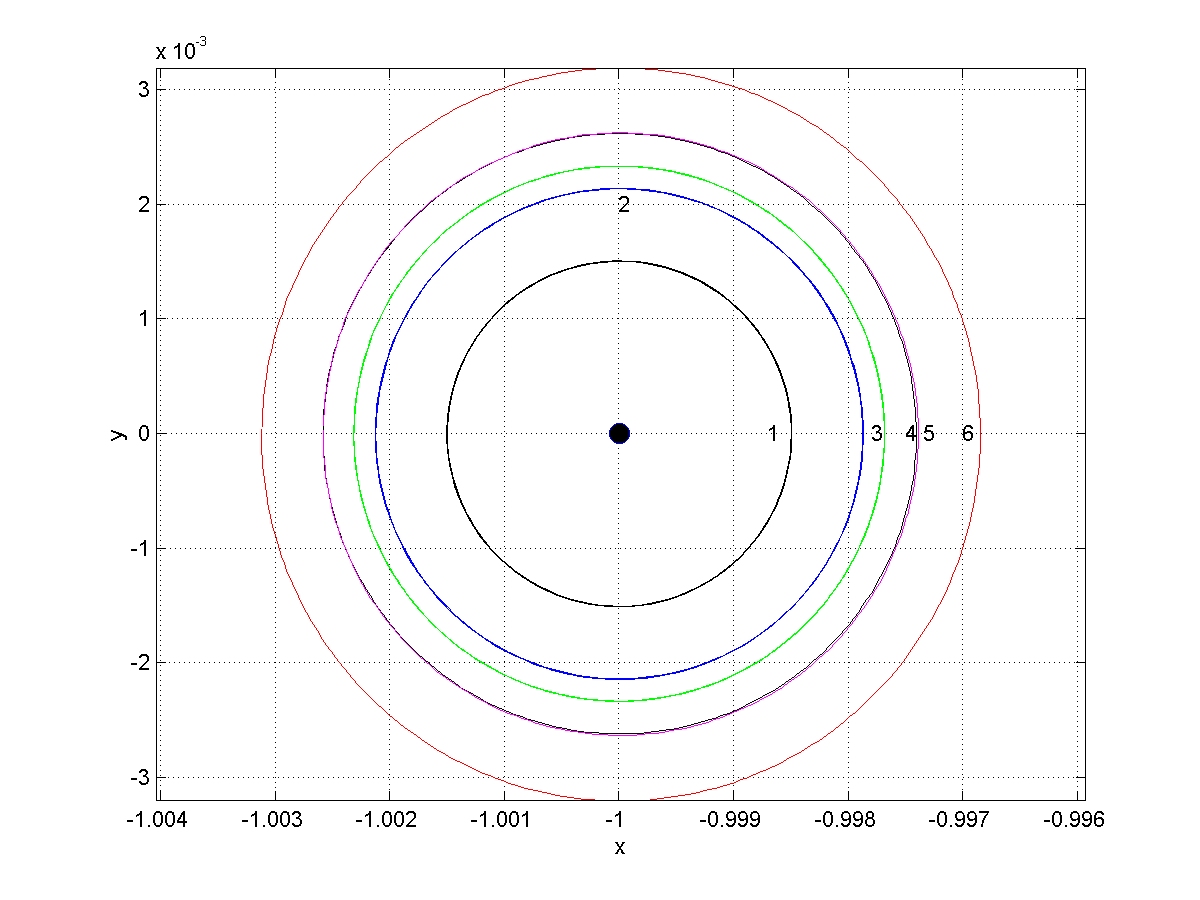}
\caption{}
\end{center}
\end{figure}

\clearpage
\begin{figure}
\begin{center}
\includegraphics[width=160mm]{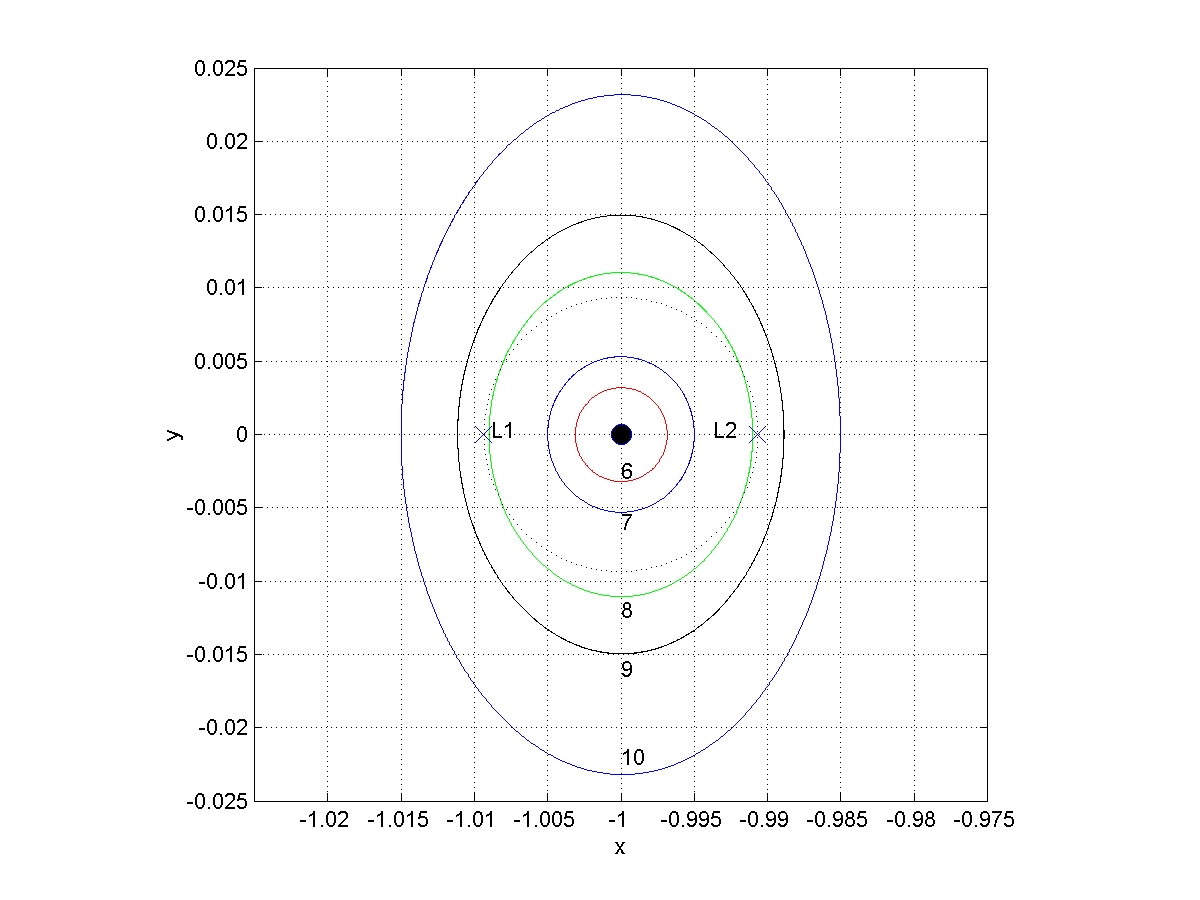}
\caption{}
\end{center}
\end{figure}

\clearpage
\begin{figure}
\begin{center}
\includegraphics[width=160mm]{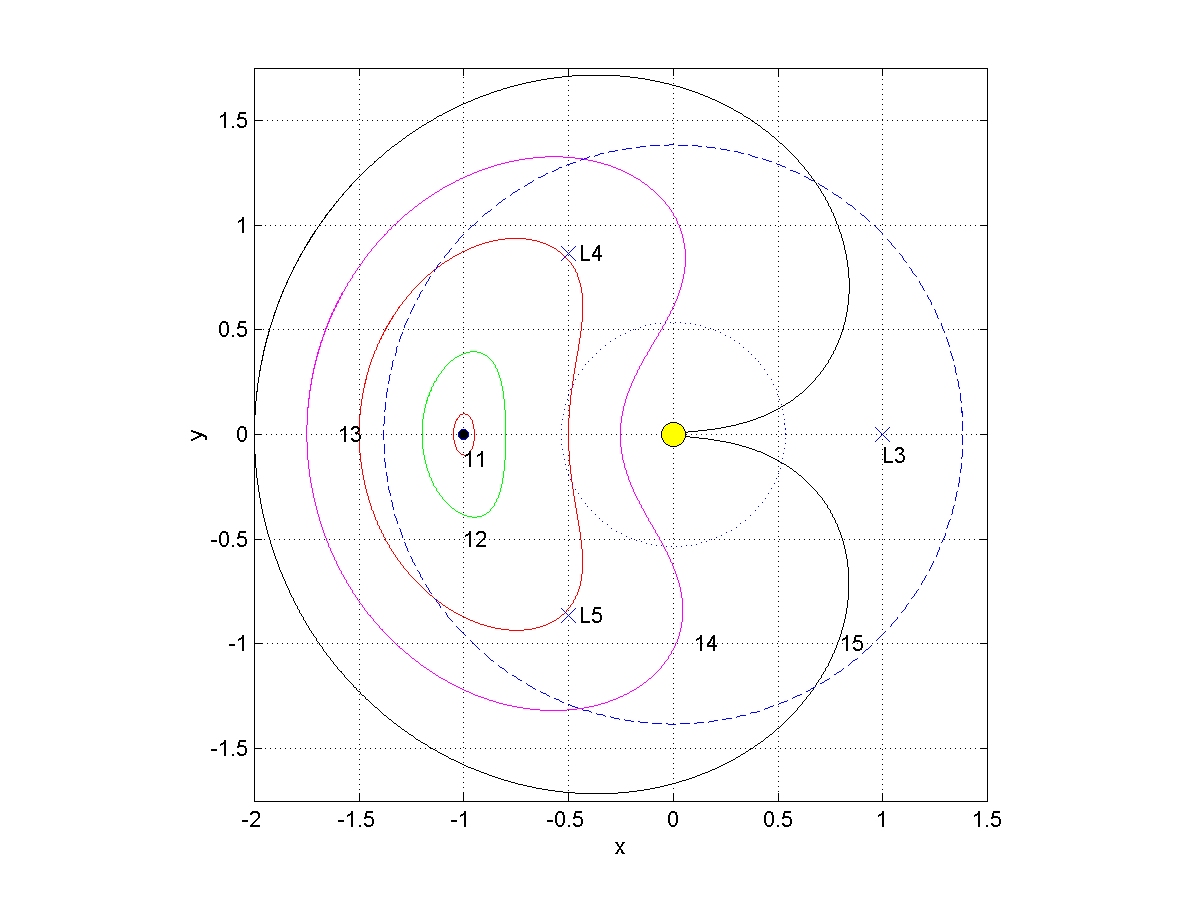}
\caption{}
\end{center}
\end{figure}

\clearpage
\begin{figure}
\begin{center}
\includegraphics[width=160mm]{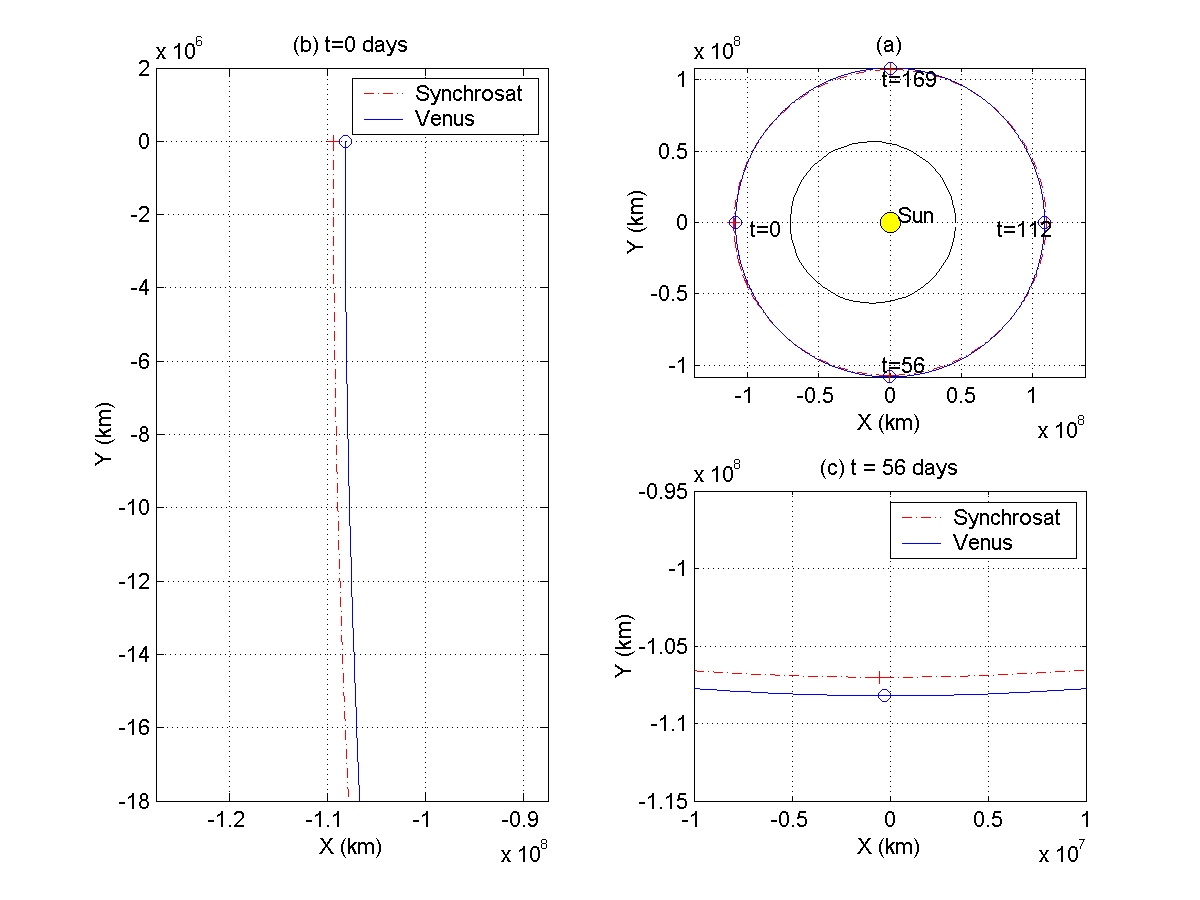}
\caption{}
\end{center}
\end{figure}

\clearpage
\begin{figure}
\begin{center}
\includegraphics[width=160mm]{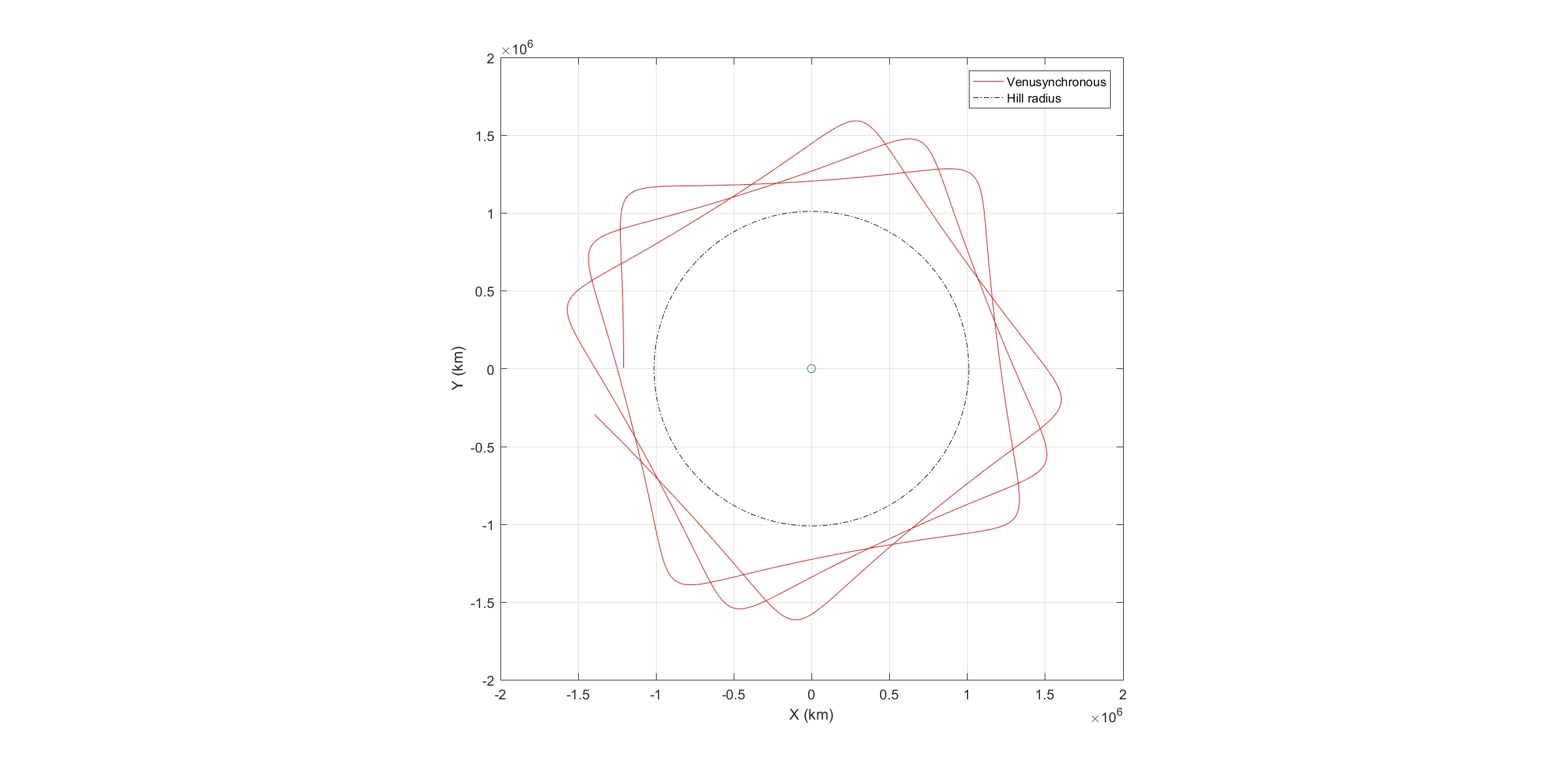}
\caption{}
\end{center}
\end{figure}

\clearpage
\begin{figure}
\begin{center}
\includegraphics[width=160mm]{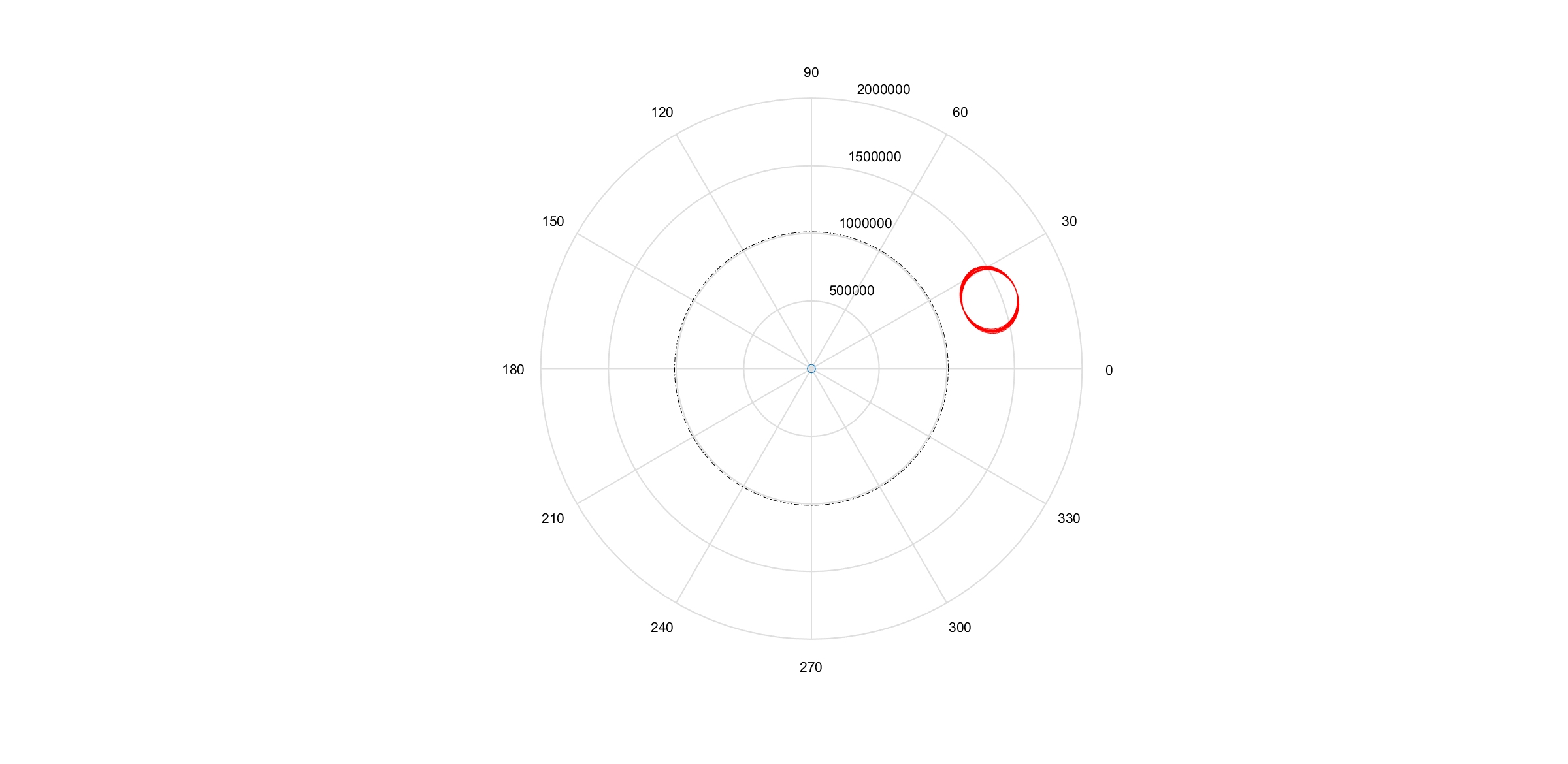}
\caption{}
\end{center}
\end{figure}

\clearpage
\begin{figure}
\begin{center}
\includegraphics[width=160mm]{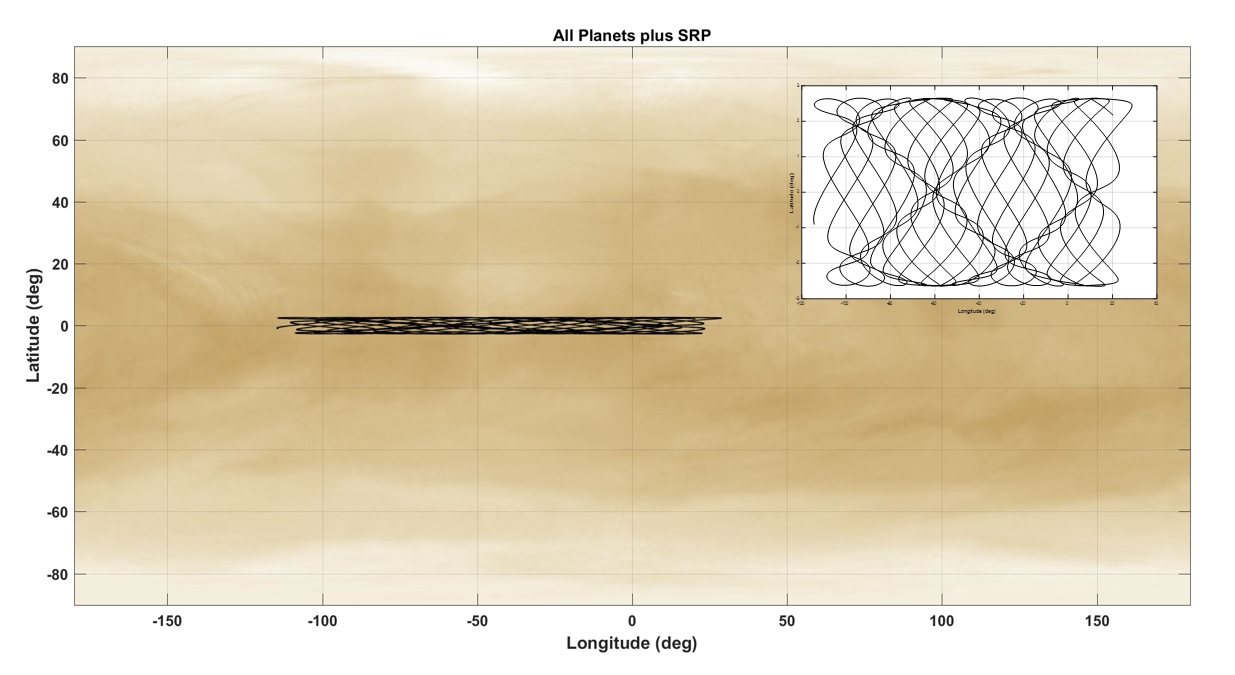}
\caption{}
\end{center}
\end{figure}

\clearpage
\begin{figure}
\begin{center}
\includegraphics[width=160mm]{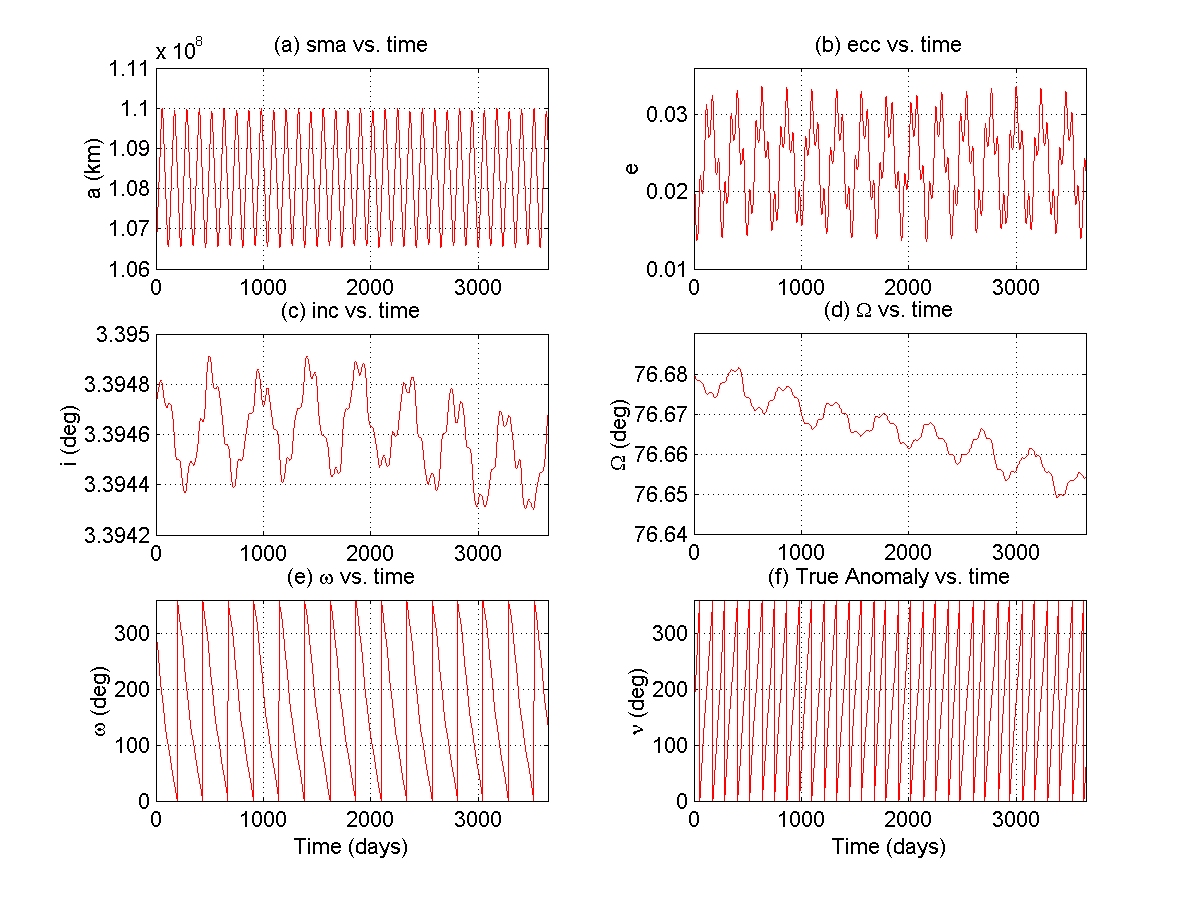}
\caption{}
\end{center}
\end{figure}

\end{document}